\documentclass[12pt,a4paper]{article}
\usepackage{amsmath}
\usepackage[dvips]{graphicx}
\input epsf
\usepackage{times}
\begin{document}
\begin{titlepage}
\title{Note on inconsistency of  unitarity saturation and maximal odderon}
\author{S.M. Troshin\footnote{e-mail: Sergey.Troshin@ihep.ru}\\[1ex]
\small  \it Institute for High Energy Physics,\\
\small  \it Protvino, Moscow Region, 142281, Russia} \normalsize
\date{}
\maketitle

\begin{abstract}
It is shown  that regime of elastic scattering with maximal odderon contribution
is not compatible with unitarity  and black disk limits saturation.
\end{abstract}
\begin{center}
\end{center}
\end{titlepage}
\setcounter{page}{2}
Studies of elastic scattering becomes a fascinating
 subject in the light  of the coming
start of data taking at the LHC
 \cite{jen}. Various theoretical schemes exist in the field.
Most of those approaches are compatible with the results obtained in axiomatic field
theory and they can also fit to the existing experimental data. However, use
 of QCD in this field is problematic due to the unsolved problem of confinement.
 Certainly, it is difficult to incorporate
all  known
 dynamical issues and limitations into a particular phenomenological model. But it  is
 also difficult to expect
 that the model inconsistent with unitarity (i.e. the one violating probability
 conservation law) would adequately  reflect  the  dynamics of hadron interaction and
 provide reliable predictions.
 To fulfill unitarity condition under a model construction of the elastic
 amplitudes, it is natural to use unitarization approaches such as eikonal or U-matrix, which consider
  amplitudes in the  impact parameter space. They automatically guarantee that elastic amplitude in the impact
  parameter representation will obey unitarity condition and the inequality $|f(s,b)|\leq 1$, in particular.

 Despite that the full implementation of unitarity is not possible at the moment (cf. e.g. \cite{land}), the
 amplitude in the impact parameter space should not exceed unity anyway.
 However, it might  be possible not the case when the amplitude is constructed in
  the $s$ and $t$ representation; it is a priori not evident that the particular form
  of the amplitude $F(s,t)$ when
   transformed into the impact parameter space  will satisfy unitarity and be less than unity.
  This is true even in the case, when the
  model leads to the predictions for observables and they  explicitly agree with axiomatic bounds, e.g.
 such as the well known Froissart-Martin bound for the total
 cross--sections. Agreement with experimental data at finite energies
 and with asymptotical bounds is not enough since  wide class of
 functional dependencies can describe experimental data well
 and have correct asymptotical behavior.
Additional cross check is  needed  to prove that the impact parameter
 amplitude, namely its real and imaginary parts are in agreement with unitarity
   at finite as well as asymptotic energies.

 The principle of maximum strength for strong interactions was proposed by Chew and Frautschi in
 \cite{chew}. It was supposed, in particular, that strong interactions will
 saturate unitarity condition at $s\to\infty$.
 However, more than three decades  ago it was assumed that maximality of the strong interactions
  strength
  would correspond to the maximally possible increase of the crossing-even and crossing-odd forward amplitudes
  (linear combinations of $pp$ and $\bar p p$ amplitudes)\cite{nicol}, which
  with account of the Phgragm\'{e}n-Lindel\"{o}f
  theorem (cf. \cite{block}), can be translated into the following simultaneous dependencies  of the
  imaginary part of the forward $pp$-scattering scattering amplitude and its real part
 \[\mbox{Im}F(s,t=0)\sim s\ln^2 s,\] \[\mbox{Re}F(s,t=0)\sim s\ln^2 s .\]
 This regime was supposed to  result from maximal odderon contribution
 and it was used to construct phenomenological
description of elastic scattering data and provide  predictions for the LHC energies
in the recent papers (cf. e.g. \cite{mart}).
 However, the amplitudes in the impact parameter space were not calculated  and therefore
 a real danger  of unitarity violation exists.
Indeed, an additional unitarity restriction exists for models
which do not suppose domination of imaginary
 part of scattering amplitude.
Unitarity condition in the impact parameter representation for the
elastic scattering amplitude can
 be rewritten in the form:
 \[
 \mbox{Im} f(s,b)[1-\mbox{Im} f(s,b)]=[\mbox{Re} f(s,b)]^2+\eta(s,b),
 \]
where $0\leq \eta(s,b)\leq 1/4$ is the contribution of inelastic
channels.
 Since \[
 0\leq\mbox{Im}(s,b)\leq 1
 \]
  we obtain that unitarity   limits the real part of scattering amplitude
 (which can be sign changing function contrary to $\mbox{Im}(s,b)$) in the following
 way
 \[
[\mbox{Re} f(s,b)]^2\leq 1/4,
\]
\[
-\frac{1}{2}\sqrt{1-4\eta(s,b)}\leq \mbox{Re} f(s,b)\leq \frac{1}{2}\sqrt{1-4\eta(s,b)}.
\]

This limitation, as it was already mentioned, is essential for the
models with odderon and is indirectly in favor of the standard
procedure of neglecting the real part of scattering amplitude
compared to its imaginary part. It  also is  evident that absolute
value of the real part and imaginary part of elastic scattering
amplitude $f(s,b)$ cannot reach their maximal values
simultaneously, moreover when $ \mbox{Im} f(s,b)\to 1$, saturating
unitarity limit at large values of s in the region $b<R(s)$, then
$\mbox{Re} f(s,b)\to 0$ in this kinematical region. It means that
$[\mbox{Re} f(s,b)]^2$ should have  a peripheral impact parameter
profile in this case.
 The same conclusion is valid when $ \mbox{Im} f(s,b)\to 1/2$,
saturating the black disk  limit at large values of s in the
region $b<R(s)$, then $\mbox{Re} f(s,b)\to 0$ because
$\eta(s,b)\to 1/4$ in this region. The above difference in the
impact parameter profiles would result in the different energy
dependencies of $ \mbox{Im} F(s,t=0)$ and $ \mbox{Re} F(s,t=0)$
bringing maximal odderon on the edge of contradiction with
unitarity (or black disk) saturation as it will be demonstrated in
the following. The
unitarity
 condition itself can  be  obeyed  by a scattering amplitude with
 maximal odderon contribution\footnote{An explicit specific
  example of
  the amplitudes was considered in \cite{gar}  using the eikonal form and
 it was shown, that in this particular case, unitarity condition is  satisfied.
 Since the eikonal complex phase is bounded at $s\to\infty$, this amplitude does not
 saturate unitarity limit but can provide  saturation of the Froissart--Martin bound at the price of
 $\sigma_{inel}(s)\to 0$ at $s\to\infty$. Real and imaginary parts of the scattering amplitude have the
 similar central impact parameter profiles in this case.}.
 We prove that saturation
 of unitarity is in contradiction with maximal odderon, i.e. if one supposes that
 the elastic unitarity limit is saturated at asymptotical energies, ( $\mbox{Im} f(s,b)$ has
 maximal value equal to
 unity at $b\leq R(s)$, where $R(s)$
 is the effective interaction radius) then there is no room for the asymptotical amplitude behavior
 corresponding to the maximal odderon contribution. It should be noted that at $s\to\infty$
 effective interaction radius has logarithmic  energy dependence
  $R(s)\sim\frac{1}{\mu}\ln s$.
 Therefore  unitarity saturation is a natural mechanism  of total cross-section growth in the form
$\sigma_{tot}(s)\sim \ln^2 s$ at $s\to\infty$ and it can be related to confinement \cite{citan}.
It should be noted that this mechanism does not suppose  that $\mbox{Re} f(s,b)$ vanish everywhere.
Then, at very high energies
 \begin{equation}\label{stot}
 \sigma_{tot}(s)=4\pi R^2(s).
 \end{equation}
On the other hand,
there is an inequality \cite{singh,logunov}
\begin{equation}\label{rho}
\rho^2 (s)+1\leq 4\pi R^2(s)\frac{\sigma_{el}(s)}{\sigma^2_{tot}(s)}.
\end{equation}
for the ratio of the real to imaginary part of the forward
elastic scattering amplitude $\rho(s)\equiv \mbox{Re} F(s,t=0)/\mbox{Im} F(s,t=0)$.
The inequality (\ref{rho}) was derived on the basis of unitarity and dispersion relations for
scattering amplitude.
From Eqs. (\ref{stot}) and (\ref{rho}) one can easily obtain that at
asymptotic energies $\rho(s)\to 0$, since saturation of unitarity
implies, that $\frac{\sigma_{el}(s)}{\sigma_{tot}(s)}\to 1$
 at $s\to \infty$ (cf. e.g. \cite{intj}). This is in an evident contradiction with the prediction
 of the maximal odderon regime where $\rho(s)\to const.\neq 0$.

 One may argue, that
 we should expect saturation of black disk limit $\mbox{Im} f(s,b)\leq 1/2$, instead of saturation of the
 unitarity limit, i.e. one would take asymptotical relation $\frac{\sigma_{el}(s)}{\sigma_{tot}(s)}\to 1/2$.
 This belief is based on the assumption of the absorptive effects domination. Such domination
 results from the use of an eikonal representation for the amplitude and leads to the Pumplin bound \cite{pump}
 $\sigma_{el}(s)\leq \frac{1}{2}\sigma_{tot}(s)$.
 In this case at very high energies
 $\sigma_{tot}(s)=2\pi R^2(s)$
 and we again arrive to the $\rho(s)\to 0$, i.e. the same contradiction
between maximal odderon and black disk saturation takes place. This result reproduces conclusion made in
\cite{fin}.

Thus, one can conclude that saturation of elastic unitarity (or
black disk) limit leaves no room for maximal odderon at asymptotics
 and it is inconsistent with this hypothesis. Our purpose was not to check consistency of all various amplitude
parameterizations (with many free parameters) based on maximal odderon contributions with
unitarity or black disk limitations, we have pursued a more modest aim, namely we have explicitly
demonstrated that such parameterizations are inconsistent with
unitarity saturation.

Our conclusion is not quite new. As it was already noted, similar
conclusion was made in \cite{fin} on the base of eikonal amplitude
unitarization. The present result was obtained in other way and
generalized for the case of unitarity saturation.
The remark of \cite{fin}  that phenomenology based on maximal odderon cannot be
excluded at finite energies on the theoretical grounds is definitely true but
 appears to have a little experimental confirmation,
the quantitative analysis of the available
experimental data \cite{block}  leads to  conclusion on the smallness of the odderon amplitudes.
It should be stressed that we supposed saturation of unitarity limitation for the impact parameter amplitude,
 but we did not suppose that the scattering
amplitude is the pure imaginary and cross--even one for all values of kinematical variables.
Those assumptions are not equivalent.
Of course, unitarity or black disk limits saturation itself does
not follow
 from axiomatic field theory, but
 we would like to note,
that it is much more natural to expect that  it could be  a manifestation of
a maximal strength of strong interaction instead of behavior of the real
part of the forward scattering amplitude in the form $\mbox{Re}F(s,t=0)\sim s\ln^2 s$
as it happens in the models incorporating the maximal odderon regime.

The author would like to thank N.E. Tyurin for the helpful
comments and discussions, J.P. Ralston for the interesting
correspondence on the role of unitarity in hadron physics, E.S. Martynov and V.A.
Petrov for the useful remarks.

\end{document}